\documentclass[prd,aps,preprint,11pt,onecolumn,nofootinbib]{revtex4-1}

\usepackage{amsfonts,amsmath,amssymb,graphicx,bm,subfigure,hyperref}

\begin{document}

\title{Study of neutrino spin oscillations in a gravitational field with a differential equations method}

\author{Mridupawan Deka$^{1}$}
\email{mpdeka@theor.jinr.ru}

\author{Maxim Dvornikov$^{1, 2}$}
\email{maxim.dvornikov@gmail.com}

\affiliation{$^{1)}$\,Bogoliubov Laboratory of Theoretical Physics, Joint Institute for Nuclear Research, Dubna, Russia \\
  $^{2)}$\,Pushkov Institute of Terrestrial Magnetism, Ionosphere and Radiowave Propagation (IZMIRAN), Troitsk, Moscow, Russia}

\begin{abstract}
  In this work, we employ Ordinary Differential Equation solution method
  to study neutrino spin oscillations in the case when they are
  gravitationally scattered off a rotating Kerr black hole. Previously, this problem involved the integral solution of the Hamilton-Jacobi equation. We analyze the consistency of these two methods.
\end{abstract}

\maketitle

\section{Introduction}
\label{sec:INTR}

In addition to having non-zero masses and flavor oscillations (see,
e.g., Refs.~\cite{NOvA:2021nfi,GiuKim07}), neutrinos are believed
to have non-zero magnetic moment~\cite{Beda:2012zz, Dvornikov:2003js,
  Broggini:2012df, Giu16}. The non-zero magnetic moment of neutrino leads to
neutrino spin oscillations due to the interactions of neutrinos with the magnetic fields~\cite{Fujikawa:1980yx}. Electroweak interaction with background matter also can contribute to the neutrino spin precession (see, e.g., ~\cite{Giunti:2014ixa,Giunti:2024gec}). Spin oscillations render the 
left handed active neutrinos to sterile right handed ones within the same
flavor. 

One of the possibilities that allows us to examine the neutrino spin
oscillations is the observation of gravitationally scattered ultra-relativistic
neutrinos off a black hole (BH) surrounded by an accretion disk. Several studies aiming at
the spin oscillations in curved spacetime within General Relativity are carried out in
Refs.~\cite{Dvo06,Dvo13,Dvo23c,Dvo23d,Dvo23a,Dvo23b,Deka:2023ljj,Deka:2025eub,Deka:2025war}.
These studies employ the quasi-classical approach for the the motion of
the spinning particles described in Ref.~\cite{PomKhr98}. In a gravitational
scattering, both ``in'' and ``out'' states of neutrinos are in
the asymptotically flat spacetime. Therefore, their spin states are well
defined.

In our previous
  works~\cite{Dvo06,Dvo13,Dvo23c,Dvo23d,Dvo23a,Dvo23b,Deka:2023ljj,
    Deka:2025eub,Deka:2025war}, where we have studied spin oscillations
  of neutrinos gravitationally scattered by BH, first, we
  reconstructed the particle trajectories using the integral solution of
  the Hamilton-Jacobi equation. It is a commonly accepted method to study
  the motion of test particles in the gravitational field of BH
  (see, e.g., Ref.~\cite{Hac19}). Then, we have used the obtained trajectories
  to study the spin evolution along each of these trajectories. The spin precession
  equation (see Eq.~\eqref{eq:nuspinrot} below) does not allow a simple
  solution in the general case\footnote{If the vector $\bm{\Omega}$ has
  three components which arbitrarily depend on time, the analytical solution of
  Eq.~\eqref{eq:nuspinrot} exists only in a few specific cases.}.
  Thus, one has to solve the spin precession equation numerically.

  To implement the numerical solution of the spin evolution equation, one can
  interpolate the obtained trajectories by smooth functions to use, e.g., the
  Runge--Kutta method with the adaptive step size. However, the interpolation
  is a computationally expensive procedure. It becomes especially noticeable
  when one deals with a great number of test particles.

  In Ref.~\cite{Deka:2025eub}, we have used the predictor-corrector methods,
  which allowed us to evaluate the right hand side of Eq.~\eqref{eq:nuspinrot}
  in several points of the grid. These methods have made it possible to increase
  the accuracy of computations. However, the spin evolution equation can become
  stiff in certain integration points.
  In these situations, it is not clear how tiny the grid
  steps have to be. In this regard, it is better to do a numerical
  check by using an alternative approach where the grid steps
  does not have to be defined \textit{a priori}.

  We also list several technical problems which one encounters while describing spin oscillation of scattered neutrinos. The evaluation of trajectories of neutrinos heavily depends
on numerical computations of (In)complete elliptic integrals of first and third kind. For this purpose, one has to set up an appropriate
grid in particle position, $r$. The choice of the grid can be appropriate for the reconstruction of the trajectories. However, it can be ambiguous for the spin evolution equation. The behavior of trajectories is quite sensitive to the grid partition near turn points in terms of $\phi(r)$ and $\theta(r)$ dependencies. Any computational errors in the trajectories can accumulate
and lead to even greater inaccurate results in the spin evolution.

It is therefore desirable to have an alternative method which is devoid of the above
issues. This alternative method shall also act as a numerical checks of our previous
results. Hence instead of solving equations that involve integrals, we shall try to
solve the corresponding differential equations using adaptive methods. The use of
adaptive method ensures that the $r$-grid will automatically be determined by the algorithm,
and it will be different for different neutrinos depending upon the gravitational
interactions. On the other hand, the spin evolution equations are also in differential
form, and they depend only on $r$ and $\theta$. So the differential equations for
particle motion and spin evolution can simultaneously be solved. Note that we do not need
$r$ dependence of the azimuthal angle, $\phi$, along the whole trajectory. We only need
the value of $\phi$ at the observer position. Therefore, we use the $r$-grid obtained
from the adaptive solution method to evaluate $\phi$. It should be noted that the reconstruction of trajectories of ultrarelativistic particles by solving the differential geodesics equations was made in Refs.~\cite{Zak91,Zak94}.

In this work,  we restrict ourselves only to the gravitational interactions. Although
the particle motion is fully described by $r$, $\theta$ and $\phi$, we determine
$r$ and $\theta$ only. We also limit ourselves to a few 
hundred randomly chosen incoming test neutrinos.
In future, we plan to do a more extensive work which will also include electromagnetic
and electroweak interactions of neutrinos. This will require introduction of a
magnetized accretion disk.

This work is organized in the following way. First, in Sec.~\ref{sec:formalism},
we briefly outline our approach, and discuss the motion and the spin evolution of
a test particle in the gravitational field of a rotating BH, respectively. The
numerical parameters are described in Sec.~\ref{sec:NUMERICAL}. The results are
then presented in Sec.~\ref{sec:RES}. Finally, we conclude our work in
Sec.~\ref{sec:CONCL}.

\section{Formalism}
\label{sec:formalism}

We suppose that a beam of incoming neutrinos, which are emitted from a source with
the coordinate, $(r,\theta,\phi)_s = (\infty,\theta,0)$, is traveling in the vicinity
of a spinning supermassive BH (SMBH). While some of them fall into the BH, the rest 
are scattered gravitationally. The gravitationally scattered neutrinos
eventually escape to infinity, and finally are observed at the position,
$(r,\theta,\phi)_{\mathrm{obs}} = (\infty,\theta_\mathrm{obs},\phi_\mathrm{obs})$.
Since we consider only gravitational interactions, we do not expect any
spin oscillations. Our goal is to study the probability distributions of spin
states of these scattered neutrinos as functions of $\theta_\mathrm{obs}$ and
$\phi_\mathrm{obs}$, and compare them with those obtained from methods involving
integrals. Below we define the trajectories and spin evolution of these neutrinos between
$(r,\theta,\phi)_s$ and $(r,\theta,\phi)_{\mathrm{obs}}$. 


We describe the spacetime of a spinning BH in Kerr metric. For a BH with the
mass $M$ and angular momentum $J$ along the $z$-axis, the metric can be written
in Boyer\---Lindquist coordinates, $x^{\mu}=(t,r,\theta,\phi)$, as,
\begin{equation}\label{eq:Kerrmetr}
  \mathrm{d}s^{2}=g_{\mu\nu}\mathrm{d}x^{\mu}\mathrm{d}x^{\nu}=
  \left(
    1-\frac{rr_{g}}{\Sigma}
  \right)
  \mathrm{d}t^{2}+2\frac{rr_{g}a\sin^{2}\theta}{\Sigma}\mathrm{d}t\mathrm{d}\phi-\frac{\Sigma}{\Delta}\mathrm{d}r^{2}-
  \Sigma\mathrm{d}\theta^{2}-\frac{\Xi}{\Sigma}\sin^{2}\theta\mathrm{d}\phi^{2},
\end{equation}
where,
\begin{equation}\label{eq:dsxi}
  \Delta=r^{2}-rr_{g}+a^{2},
  \quad
  \Sigma=r^{2}+a^{2}\cos^{2}\theta,
  \quad
  \Xi=
  \left(
    r^{2}+a^{2}
  \right)
  \Sigma+rr_{g}a^{2}\sin^{2}\theta.
\end{equation}
Here, $r_g = 2M$ is the Schwarzschild radius and $J = a M$, where $0 < a < M$ .

For the first time, the Hamilton-Jacobi equation for a test particle in the Kerr metric in Eq.~\eqref{eq:Kerrmetr} was solved in Ref.~\cite{Carter:1968rr}. It was found that the geodesic motion of an ultra-relativistic test particle in Kerr metric has three
constants of motion: the particle energy, $E$, its angular momentum, $L$, and the
Carter constant, $Q$. Since we are considering scattering only, therefore $Q>0$. The recent description of the particle motion in Kerr metric was made in Ref.~\cite{GraLupStr18}.
The trajectory of an ultra-relativistic neutrino in the presence of the
gravitational field of a spinning SMBH can be written
as~\cite{Carter:1968rr,GraLupStr18},
\begin{align}
  & \int\frac{\mathrm{d}r}{\pm\sqrt{R}}=\int\frac{\mathrm{d}\theta}{\pm\sqrt{\Theta}},
  \label{eq:trajth}
  \\
  & \phi = a\int\frac{\mathrm{d}r}{\pm\Delta\sqrt{R}}[(r^{2}+a^{2})E-aL]+
  \int\frac{\mathrm{d}\theta}{\pm\sqrt{\Theta}}
  \left[
    \frac{L}{\sin^{2}\theta}-aE
  \right].
  \label{eq:trajphi}
\end{align}
where $R$ and $\Theta$ potentials are defined as,
\begin{align}
  R(r) = & [(r^{2}+a^{2})E-aL]^{2}-\Delta[Q+(L-aE)^{2}],\notag
  \\
  \Theta(\theta)= & Q + \cos^{2}\theta
  \left(
    a^{2} E^{2} - \frac{L^{2}}{\sin^{2}\theta} \notag
  \right). \notag
\end{align}
The choices that determine the $\pm{}$ signs for $\sqrt{R}$ and $\sqrt{\Theta}$ in
Eqs.~\eqref{eq:trajth} and~\eqref{eq:trajphi} have been discussed in detail in
Ref.~\cite{Deka:2025eub}.

It is reasonable to separate all the incoming neutrinos by the sign of $\dot{\theta}$ at $t\to -\infty$~\cite{Deka:2025eub}. If $\dot{\theta}_{-\infty} < 0$, we call such a neutrino as an ``Upper'' one. Otherwise, one deals with a ``Lower'' neutrino. If the angle of incidence,
  $\theta_i = \pi/2$, the Upper and Lower incoming neutrinos propagate above and below the equatorial plane at $t\to -\infty$, respectively.

By defining the dimensionless variables,
$r = xr_{g},\, L = yr_{g}E,\, Q = wr_{g}^{2}E^{2},\, a = zr_{g},\,
\mathfrak{\tilde t} = \cos\theta$, we can rewrite the Eq.~\eqref{eq:trajth} as,
\begin{align}
  & \frac{1}{z}\frac{\mathrm{d}\mathfrak{\tilde t}} {\mathrm{d}x}
  =
  \frac{\pm\sqrt{\Theta(\mathfrak{\tilde t})}} {\pm\sqrt{R(x)}},
  \label{eq:trajth_1}
\end{align}
where
\begin{align}
  \label{eq:RTh}
  R(x) =&
  \left(
    x^{2}+z^{2}-yz
  \right)^{2}-(x^{2}-x+z^{2})
  \left[
    w+
    \left(
    z-y
    \right)^{2}
    \right], \\
  \Theta(\mathfrak{\tilde t}) =&(\mathfrak{\tilde t}_{-}^{2}+\mathfrak{\tilde t}^{2})
  (\mathfrak{\tilde t}_{+}^{2}-\mathfrak{\tilde t}^{2}),\\
  \mathfrak{\tilde t}_{\pm}^{2} =&
  \frac{1}{2z^{2}}\left[\sqrt{(z^{2}-y^{2}-w)^{2}+4z^{2}w}\pm(z^{2}-y^{2}-w)\right],
\end{align}
with $\mathfrak{\tilde t}_{\pm}^2>0$. The value of the variable, $\mathfrak{\tilde t}$, as a
function of $x$ in Eq.~\eqref{eq:trajth_1} oscillates between $\pm \mathfrak{\tilde t}_{+}$,
$- \mathfrak{\tilde t}_{+} < \mathfrak{\tilde t} < + \mathfrak{\tilde t}_{+}$. The
  lower limit of $x$ is set at $x = (x_{\mathrm{tp}} + \epsilon)$ where
  $\epsilon \rightarrow 0$. $x_{\mathrm{tp}}$ is the turn point which is the maximal real
  root of the equation $R(x) = 0$ with $R(x)$ being given in Eq.~\eqref{eq:RTh}. This is
  the minimal distance to the BH center.  Note that we add $\epsilon$ to $x_{\mathrm{tp}}$
  for the lower limit of $x$. This is to avoid singularity at $x = x_{\mathrm{tp}}$. The
  addition of $\epsilon$ works in our case, because it has been shown in
  Ref.~\cite{GraLupStr18} that the integral is convergent at the
  leading order in the $\epsilon$-expansion.

The following parametric equations define the BH shadow curve between the captured
and escaped neutrinos~\cite{Deka:2025eub,GraLupStr18},
\begin{eqnarray}
  \label{eq:y}
  y &=& - \frac{1}{z (2x - 1)}
  \left[x^2 (2x -3) + z^2 (2x + 1)\right],\\
  \label{eq:w}
  w &=& \frac{x^3}{z^2 (2x - 1)^2}
  \left[8z^2 - x (2x - 3)^2\right],\\
  \label{eq:rtilde}
  x_{\pm} &=& 1 + \cos\left[\frac{2}{3} \arccos(\pm 2z)\right],
  \hspace{2mm}
  x_{-} < x < x_{+}.
\end{eqnarray}
Equations~\eqref{eq:y}-\eqref{eq:rtilde} guarantee that $\Theta(\mathfrak{\tilde t})>0$
in Eq.~\eqref{eq:trajth_1} if $\mathfrak{\tilde t}_i = \cos\theta_i = 0$,
i.e. when the incoming flux is parallel to the equatorial plane. The curve defined by Eqs.~\eqref{eq:y}-\eqref{eq:rtilde} separates the values of the integrals of motion corresponding to particles falling into BH and being scattered off~\cite{Zak86}. If one rewrites the curve given by Eqs.~\eqref{eq:y}-\eqref{eq:rtilde} in terms of the screen coordinates, one arrives at the concept of the BH shadow. BH shadows for different observer positions are presented in Ref.~\cite{Zak05}. The recent reviews on the studies of BH shadows are presented in Refs.~\cite{Zak24,Zak25}. BH shadows which appear in alternative theories of gravity are considered in Ref.~\cite{Zak25b}.

If $\mathfrak{\tilde t}_i \neq 0$, we should demand that
$|\mathfrak{\tilde t}_{i}|<\mathfrak{\tilde t}_{+}$ to have
$\Theta(\mathfrak{\tilde t})>0$ for any $\mathfrak{\tilde t}$ since the integration
variable in Eq.~\eqref{eq:trajth_1} covers the segment
$\mathfrak{\tilde t}_i < \mathfrak{\tilde t} < \mathfrak{\tilde t}_+$. Based on
Eq.~\eqref{eq:RTh}, the condition $|\mathfrak{\tilde t}_i| < \mathfrak{\tilde t}_+$
can be rewritten as,
\begin{eqnarray}
  \label{eq:w_low_bound}
  w &>& - z^2 \mathfrak{\tilde t}_i^2 + y^2 \frac{\mathfrak{\tilde t}^2_i}{1 - \mathfrak{\tilde t}^2_i},
\end{eqnarray}
which should be considered along with Eqs.~\eqref{eq:y}-\eqref{eq:rtilde}.

The polarization of a neutrino can be described by an invariant three vector
$\bm{\zeta}$ in the rest frame with respect to a locally Minkowskian frame.
The evolution of the neutrino polarization vector obeys,
\begin{equation}\label{eq:nuspinrot}
  \frac{\mathrm{d}\bm{\bm{\zeta}}}{\mathrm{d}t}=2(\bm{\bm{\zeta}}\times\bm{\bm{\Omega}}),
\end{equation}
    The explicit form of ${\bm{\Omega}}\equiv{\bm{\Omega}}_{\mathrm{g}}$ in the presence of the only gravitational interaction, which has been derived in
      Refs.~\cite{Dvo23a,Dvo23b,Deka:2025eub}, reads
\begin{equation}\label{eq:vectG}
  \bm{\bm{\Omega}}_g=
    \frac{1}{2U^{t}}
    \left[
      \mathbf{b}_{g}+\frac{1}{1+u^{0}}
      \left(
        \mathbf{e}_{g}\times\mathbf{u}
      \right)
    \right].
\end{equation}
Here $u^{a}=(u^{0},\mathbf{u})=e_{\ \mu}^{a}U^{\mu}$, where the vierbein vectors,
$e_{\ \mu}^{a}$, transform the metric in Eq.~\eqref{eq:Kerrmetr} to the Minkowskian
one, and $U^{\mu}=(U^{t},U^{r},U^{\theta},U^{\phi})$
is the four velocity of a neutrino in the world coordinates $x^\mu$. The quantities,
$\mathbf{e}_{g}$ and $\mathbf{b}_{g}$, are the components of the tensor,
$G_{ab}=(\mathbf{e}_{g},\mathbf{b}_{g})=\gamma_{abc}u^{c}$, where 
$\gamma_{abc}=\eta_{ad}e_{\ \mu;\nu}^{d}e_{b}^{\ \mu}e_{c}^{\ \nu}$
are the Ricci rotation coefficients. The semicolon stands for the covariant
derivative.

The invariant spin vector, $\bm{\zeta}$, defines the spin density matrix,
  $\rho = \tfrac{1}{2}[1+(\bm{\bm{\sigma}}\bm{\zeta})]$, which obeys the quantum
  Liouville -- von Neumann equation. Here
  $\bm{\sigma} = (\sigma_{1}, \sigma_{2},\sigma_{3})$ are the Pauli matrices.
  If a neutrino is in a pure spin state, we can consider an effective wave
  function instead of the density matrix. This effective wave function $\psi$,
  which describes the particle polarization, has been shown in
  Ref.~\cite{LobPav99} (see also Refs.~\cite{Dvo23a,Dvo23b,Deka:2025eub}) to obey
  the effective Schr\"odinger equation,
\begin{equation}\label{eq:Schreq}
  \mathrm{i}\frac{\mathrm{d}\psi}{\mathrm{d}x}= \hat{H}_{x}\psi,
\end{equation}
where
\begin{equation}
  \hat{H}_{x}= -\mathcal{U}_{2}(\bm{\bm{\sigma}}\cdot\bm{\bm{\Omega}}_{x})\mathcal{U}_{2}^{\dagger},
  \quad
  \bm{\bm{\Omega}}_{x} =  r_{g}\bm{\bm{\Omega}}_g\frac{\mathrm{d}t}{\mathrm{d}r},
  \quad
  \mathcal{U}_{2}=\exp(\mathrm{i}\pi\sigma_{2}/4).
\end{equation}
Therefore, we shall numerically solve Eq.~\eqref{eq:Schreq} instead of dealing with the precession Eq.~\eqref{eq:nuspinrot}.

The Hamiltonian $\hat{H}_{x}$ in Eq.~\eqref{eq:Schreq} is the function of $x$ only through the dependence of
$\theta(x)$  obtained from Eq.~\eqref{eq:trajth_1}. The initial
condition has the form, $\psi_{-\infty}^{\mathrm{T}}=(1,0)$, which means all incoming
neutrinos are left polarized. The solution of Eq.~\eqref{eq:Schreq} provides the
polarization of a scattered neutrino in the form,
$\psi_{+\infty}^{\mathrm{T}}=(\psi_{+\infty}^{(\mathrm{R})},\psi_{+\infty}^{(\mathrm{L})})$.
The probability that a neutrino remains left polarized at the observer position,
is $P_{\mathrm{LL}}=|\psi_{+\infty}^{(\mathrm{L})}|^{2}$.

An outgoing neutrino is supposed to be detected by a terrestrial neutrino telescope. The neutrino detector is based on the standard model physics, where only the left handed neutrinos are active. That is why the neutrino flux, observed in a neutrino telescope, is $\propto P_{\mathrm{LL}}$.

Thus, to summarize, in order to find $P_{\mathrm{LL}}$ we simultaneously solve two differential Eqs.~\eqref{eq:trajth_1} and~\eqref{eq:Schreq}. For this purpose, we use the adaptive Runge-Kutta-Fehlberg $7(8)$ method. 

We do not need $x$ dependence of the azimuthal angle $\phi$ along the whole
trajectory. We still use the explicit forms of $\phi$ for incoming and outgoing
neutrinos for both Upper and Lower cases given in Ref.~\cite{Deka:2025eub}. For
integration, we use the $x$-grid obtained from the adaptive solution method
applied to
Eqs.~\eqref{eq:trajth_1} and~\eqref{eq:Schreq}, as well as the number of turns in order to evaluate $\phi$.

Note that $\phi_\mathrm{obs} = \phi_{\mathrm{in}} + \phi_{\mathrm{out}}$. Also,
a neutrino can make multiple revolutions around the BH. This results in the
azimuthal angle, $\phi$, being greater than $2\pi$. One should account for
it in the final analysis.

\section{Parameters}
\label{sec:NUMERICAL}

In our numerical simulations, we assume that the incoming flux is uniform, with all the initial neutrino velocities being parallel. We parameterize each incoming trajectory in terms of two dimensionless numbers, $y$ and $w$, which remain constant in the subsequent particle motion. The ranges of these parameters are $-10 < y < 10$ and $0 < w < 10$. After the formation of the beam of incoming neutrinos with these $y$ and $w$, we discard the particles falling to the shadow area in Eqs~\eqref{eq:y}-\eqref{eq:rtilde}. Equation~\eqref{eq:w_low_bound} is also accounted for if one considers $\theta_i \neq \pi/2$. Therefore we guarantee that all considered particles undergo scattering only. Then, the numerical integration of Eqs.~\eqref{eq:trajth_1} and~\eqref{eq:Schreq} is carried out in the segments $x_\mathrm{max}>x > x_\mathrm{tp}$ and $x_\mathrm{tp} < x < x_\mathrm{max}$ with $x_\mathrm{max}$ is set to be equal to 100.

We consider two different spins of SMBH, namely $a\,=\,2\times 10^{-2} M$ and $0.98 M$.
We consider one angle of incidence of neutrinos, $\theta_i = 89.427^\circ$
for each BH spin. The number of randomly chosen test neutrinos we
use in each case of BH spin is more than $200$.

\section{Results}
\label{sec:RES}

\begin{figure}[htbp]
\centering
\subfigure[]
 {\label{fig:pll_z0.01_ti0.01_above}
   \includegraphics[width=0.47\hsize]{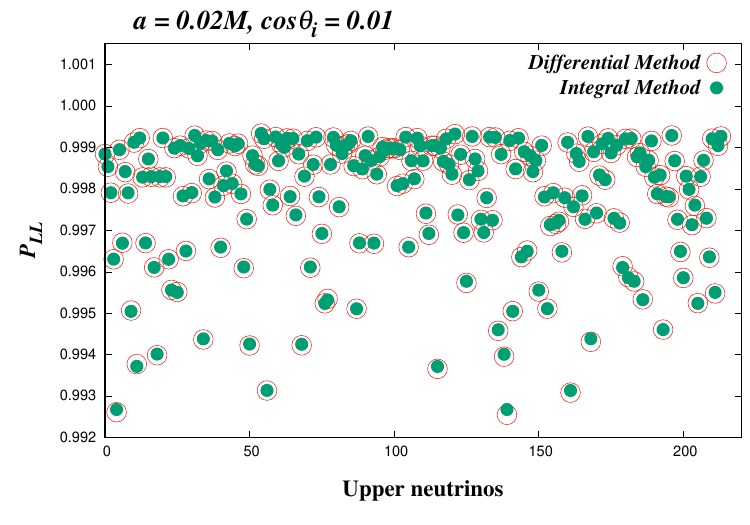}
 }
 \hspace{0mm}
 \subfigure[]
  {\label{fig:pll_z0.01_ti0.01_below}
    \includegraphics[width=0.47\hsize]{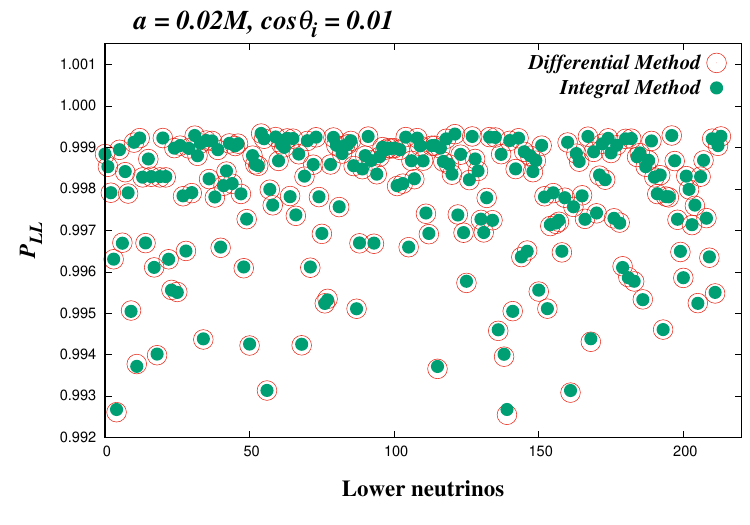}
  }
\subfigure[]
 {\label{fig:pll_z0.49_ti0.01_above}
   \includegraphics[width=0.47\hsize]{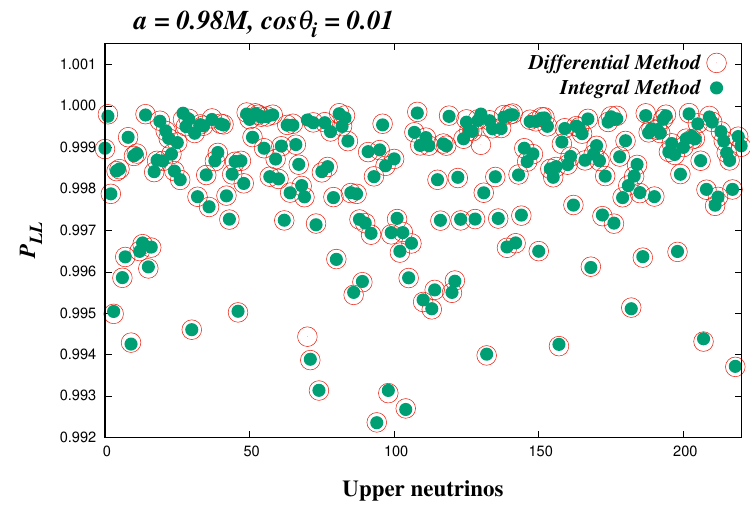}
 }
 \hspace{0mm}
 \subfigure[]
  {\label{fig:pll_z0.49_ti0.01_below}
    \includegraphics[width=0.47\hsize]{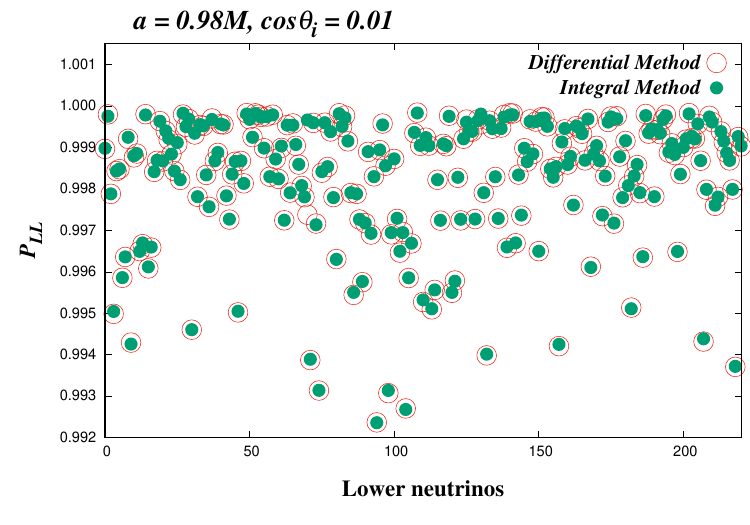}
  }   
  \caption{Comparison of $P_{\mathrm{LL}}$'s obtained from integral and differential methods.
    (a) Upper neutrinos for $a = 0.02 M$, $\cos\theta_i = 0.01$,
    (b) Lower neutrinos for $a = 0.02 M$, $\cos\theta_i = 0.01$,
    (c) Upper neutrinos for $a = 0.98 M$, $\cos\theta_i = 0.01$,
    and
    (d) Lower neutrinos for $a = 0.98 M$, $\cos\theta_i = 0.01$.}
  \label{fig:pll_comparison}
\end{figure}

In Fig.~\ref{fig:pll_comparison}, we make comparisons between $P_{\mathrm{LL}}$'s
obtained from both the integral and differential methods. The BH spin we consider
are $a\,=\,2\times 10^{-2} M$ and $0.98 M$ with the angle of of incidence of neutrinos being $89.427^\circ$. We see from these figures that both the methods produce
the similar values for $P_{\mathrm{LL}}$ for each neutrino.

In our analysis, we account for only the gravitational field contribution to the neutrino spin evolution in Eq.~\eqref{eq:vectG}. In this situation, we obtain that the survival probability for spin oscillations, $P_\mathrm{LL}$, almost equals to one. This result is obtained in both methods: the integral and differential equations one; cf. Fig.~\ref{fig:pll_comparison}. It means that no spin oscillations occur in the neutrino scattering under the influence of only the gravitational field.

One can claim that the spin-flip of ultrarelativistic neutrinos is absent in case of the scattering off a non-rotating BH~\cite{DolDorLas06}. If one considers the motion in the equatorial plane of a rotating BH, the spin-flip was shown in Ref.~\cite{Dvo21} to be absent. The suppression of spin oscillations of ultrarelativistic neutrinos in the weak gravitational lensing was predicted in Ref.~\cite{LamPapPunSca05}. However, to the best of our knowledge, the general analytical proof of this theorem does not exist. One can say that, in the present work, we check this kind of statement numerically.

Neutrino interaction with a magnetic field in an accretion disk, surrounding BH, was shown, e.g., in Ref.~\cite{Deka:2025eub} to cause spin oscillations. In this situation, one obtains $P_\mathrm{LL}<1$. For a neutrino helicity to flip in the gravitational scattering, one should have the component of the magnetic field to be transverse with respect to the neutrino velocity. We demonstrated in Ref.~\cite{Deka:2025eub} that this situation occurs if one deals with a magnetized accretion disk having a toroidal field component.

\section{Conclusion}
\label{sec:CONCL}

Our preliminary study has been very encouraging. Both the integral and
differential equations methods produce consistent results
when we consider only the gravitational interaction without an accretion
  disk. In our next  step, we plan to expand our study to
include the accretion disk so that we can take into the account the
electroweak and electromagnetic interactions in Eq.~\eqref{eq:Schreq} in
  addition to gravitational interaction. This will allow us to compare the
  results obtained by differential equations method with those in
  Ref.\cite{Deka:2025eub}, where we used the integral method.

Lastly, we mention that we are developing the numerical methods in the current stage of our research. Nevertheless, we may suggest a possible effect of neutrino spin oscillations in an observation of astrophysical neutrinos emitted, e.g., in by a core-collapsing supernova (SN) in our Galaxy. Suppose that SN, the Earth, and the SMBH in the galactic center, which neutrinos gravitationally interact with, are almost along the same line. In Ref.~\cite{Dvo23c} we estimated that the flux of scattered neutrinos can be significantly suppressed because of the neutrino spin-flip. Future neutrino telescopes like JUNO~\cite{JUNO:2015zny} and Hyper-Kamiokande~\cite{Hyper-Kamiokande:2016srs} can detect up to $\lesssim 10^5$ events if a SN explosion happens in our Galaxy. Based on our estimate in Ref.~\cite{Dvo23c}, we expect that the sensitivity of the neutrino telescopes is enough to detect reduction of the flux of scattered particles. More detailed astrophysical applications will be developed in our forthcoming works.


\begin{center}{ACKNOWLEDGMENTS}\end{center}

  All our numerical computations have been performed at Govorun super-cluster at
  Joint Institute for Nuclear Research, Dubna.

  \begin{center}{FUNDING}\end{center}
  
  The research is supported by the Joint Institute for Nuclear Research and
  the government assignment of Pushkov Institute of Terrestrial Magnetism, Ionosphere and Radio Wave
  Propagation Russian Academy of Sciences.
  
  \begin{center}{CONFLICT OF INTEREST}\end{center}
  
  The authors declare that there is no conflict of interest.

\end{document}